\documentclass{PoS}

\usepackage{amsmath,amssymb}
\usepackage{dsfont}
\usepackage{slashed}
\usepackage{verbatim}    
\usepackage{graphicx} 
 
\newcommand{\be}{\begin{equation}}
\newcommand{\ee}{\end{equation}}
\newcommand{\bea}{\begin{eqnarray}} 
\newcommand{\eea}{\end{eqnarray}}

\newcommand{\expv}[1]{\left \langle #1 \right \rangle}
\newcommand{\TAU}{\tau_f}

\newcommand{\pslash}{{\not{\hspace{-0.001cm}p}}}  

\def\Zq{Z_{\rm q}}
\def\Zm{Z_{\rm m}}

\def\ZS{Z_{\rm S}}
\def\ZP{Z_{\rm P}}
\def\ZV{Z_{\rm V}}
\def\ZA{Z_{\rm A}}
\def\ZT{Z_{\rm T}}
\def\openone{\leavevmode\hbox{\small1\kern-6.8pt\normalsize1}}

\def\chiF{\chi}

\def\MSBar{\overline{\rm MS}}

\def\lsim{\mathrel{\rlap{\lower4pt\hbox{\hskip1pt$\sim$}}
    \raise1pt\hbox{$<$}}}                
\def\gsim{\mathrel{\rlap{\lower4pt\hbox{\hskip1pt$\sim$}}
    \raise1pt\hbox{$>$}}}                

\newcommand{\Dlr}{\buildrel \leftrightarrow \over D\raise-1pt\hbox{}}
\newcommand{\Dl}{\buildrel \leftarrow \over D\raise-1pt\hbox{}}
\newcommand{\Dr}{\buildrel \rightarrow \over D\raise-1pt\hbox{}}

\title{Perturbative renormalization of staggered fermion operators with stout improvement: 
Application to the magnetic susceptibility of QCD}

\ShortTitle{Perturbative renormalization: Application to the magnetic susceptibility of QCD}

\author{G.~S.~Bali$^a$, F.~Bruckmann$^a$, M.~Constantinou$^b$, \speaker{M.~Costa}$^{\,b}$, G.~Endr\H{o}di$^a$, S.~D.~Katz$^c$,
 H.~Panagopoulos$^b$, A.~Sch\"afer$^a$\\
\llap{$^a$}Institute for Theoretical Physics, Universit\"at Regensburg, 
D-93040 Regensburg, Germany\\
\llap{$^b$}Department of Physics, University of Cyprus, Nicosia, 
CY-1678, Cyprus\\
\llap{$^c$}Institute for Theoretical Physics, E\"otv\"os University, H-1117, Budapest, Hungary\\
\\
{\rm E-mail}: \email{gunnar.bali@ur.de}, \email{Falk.Bruckmann@physik.uni-regensburg.de}, \email{constantinou.martha@ucy.ac.cy}, \email{kosta.marios@ucy.ac.cy}, \email{Gergely.Endrodi@physik.uni-regensburg.de}, \email{katz@bodri.elte.hu}, \email{panagopoulos.haris@ucy.ac.cy}, \email{Andreas.Schaefer@physik.uni-regensburg.de}}

\abstract{We calculate the fermion propagator and the quark-antiquark Green's functions for a
complete set of ultralocal fermion
bilinears,  ${{\cal O}_\Gamma}$ [$\Gamma$: scalar (S), pseudoscalar (P), vector
(V), axial (A) and tensor (T)], using perturbation theory up to one-loop and to lowest order in
the lattice spacing. We employ the staggered action for fermions and the Symanzik
Improved action for gluons.
From our calculations we determine  the renormalization
functions for the quark field and for all ultralocal taste-singlet bilinear operators.
The novel aspect of our calculations is that the gluon links which appear both in the
fermion action and in the definition of the bilinears have been improved by applying
a stout smearing procedure up to two times, iteratively. 
Compared to most other improved formulations of staggered fermions,
the above action, as well as the HISQ action, lead to
smaller taste violating effects~\cite{Aoki, Borsanyi:2010aa, Borsanyi:2010ab, Bazavov:2012}.
The renormalization functions are presented in the RI$'$ scheme; the dependence on all
stout parameters, as well as on the coupling constant, the number of colors, the lattice
spacing, the gauge fixing parameter and the renormalization scale, is shown explicitly.

We apply our results to a nonperturbative study of the magnetic susceptibility of QCD at 
zero and finite temperature. In particular, we evaluate the ``tensor coefficient'', $\tau$,
which is relevant to the anomalous magnetic moment of the muon.}

\FullConference{31st International Symposium on Lattice Field Theory LATTICE 2013\\
                 July 29 - August 3, 2013\\
                 Mainz, Germany}

\begin{document}

\section{Introduction}

Renormalization functions (RFs) are necessary ingredients in the prediction of physical probability amplitudes 
from lattice matrix elements of operators. They relate observables computed on finite 
lattices to their continuum counterparts in specific renormalization schemes.
A set of operators which are of particular interest 
are fermion bilinears which are widely employed in numerical simulations 
of Quantum Chromodynamics on the Lattice.

The quark-antiquark Green's functions of ``ultralocal'' fermion bilinears ${{\cal O}_\Gamma}$
[$\Gamma$: scalar (S), pseudoscalar (P), vector (V), axial (A) and tensor (T)], are calculated
perturbatively to one-loop order. We use massive staggered 
fermions and in the gluon sector we employ the Symanzik Improved gauge action for different 
sets of values of the Symanzik coefficients. The gluon links which appear both in the fermion 
action and in the definition of the bilinears have been improved by applying a
stout smearing procedure~\cite{Morningstar:2003gk} up to two times, iteratively. 
We implement the RI$'$ renormalization scheme, where the renormalization 
is determined by comparing the tree-level values of the quark-antiquark Green's 
functions of the operators with the corresponding values beyond tree-level. 
Some of the first perturbative results regarding staggered operators and improvements in the action 
can be found in Refs.~\cite{Patel:1992vu,Ishizuka:1993fs,Lee:2002}. 

We apply our perturbative results to a study of the response of the QCD vacuum to an 
external magnetic field, at zero and finite temperature. 
Magnetic fields probe the QCD vacuum in several ways, by affecting its fundamental 
properties like chiral symmetry breaking and restoration, the phase diagram, 
as well as the vacuum polarization. 
Here we focus on quark condensates, relevant for various phenomenological applications. 

The details of our work, along with a longer list of references, 
can be found in Refs.~\cite{Bali:2012jv,Haris:2013}.

\section{Perturbative Renormalization functions}

RFs, for operators and action parameters,
relate bare quantities, regularized on the lattice, to their
renormalized continuum counterparts: 
\be
\psi_{\rm renorm} = \Zq^{\frac{1}{2}} \psi_{\rm bare}\,,\qquad
m_{\rm renorm} = \Zm m_{\rm bare}\,,\qquad
{\cal O}^\Gamma_{\rm renorm} = Z_{{\cal O}_\Gamma} {\cal O}^\Gamma_{\rm bare}\,.
\ee
We present the RFs, in the
RI$'$ scheme, of the quark field ($\Zq$), the fermion mass ($\Zm$)
and the taste-singlet quark bilinears: scalar, pseudoscalar, vector, axial, tensor ($\ZS$,\,$\ZP$,\,$\ZV$,\,$\ZA$,\,$\ZT$).

\begin{figure}[h!]
\centering
\includegraphics[scale=0.5]{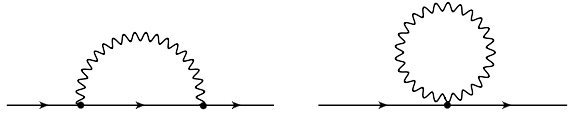}
\caption{One-loop diagrams contributing to the
fermion propagator. Wavy (solid) lines represent gluons (fermions).}
\label{figprop1}
\end{figure}

\begin{figure}[h!]
\centering
\includegraphics[scale=0.5]{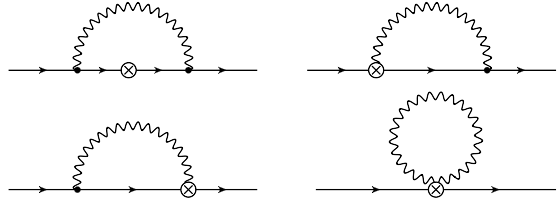}
\caption{One-loop diagrams contributing to the quark-antiquark Green's functions of the
bilinear operators. A wavy (solid) line represents gluons (fermions).
A cross denotes an operator insertion.}
\label{figbil2}
\end{figure}

The one-loop one-particle irreducible (1PI) Feynman diagrams that enter the calculation of the quark-antiquark 
amputated Green's function (inverse propagator) $S^{-1}_{1-loop}$, 
are illustrated in Fig.~\ref{figprop1}. 
For the algebraic operations involved in evaluating Feynman
diagrams, we make use of our symbolic package in Mathematica.
A brief description of the procedure for the computation of a Feynman diagram
can be found in Ref.~\cite{Constantinou:2009tr}.
The 1PI Feynman diagrams that enter in the
calculation of the two-point Green's functions of the operators, are shown in
Fig.~\ref{figbil2}, and include up to two-gluon vertices extracted from
the operator (the cross in the diagrams). The appearance of gluon
lines attached to the operators stems from the fact that the definitions of
these operators in the staggered formulation contain products of gauge links~\cite{Patel:1992vu}.

We have computed the one-loop inverse fermion propagator, $S^{-1}(p)$, for general values of: the gauge
parameter $\alpha$, the stout smearing parameters of the action (for the first and second stout 
iteration respectively) $\omega_{A_1},\,\omega_{A_2}$, the Lagrangian mass $m$, the number of colors
$N_c$ and the external momenta $p_1$, $p_2$.                 
\bea
&&S^{-1}_{1-loop} =
\frac{i}{a} \sin(a\,p_{1\rho}) \delta(p_1- p_2+\displaystyle\frac{\pi\bar\rho}{a})\nonumber\\
&&\quad\Bigg[
        1 + \frac{g^2\,C_F}{16 \pi^2}\,\Bigg[( e_1  
- \alpha\,\left(-4.79201+ \log\left(a^2 m^2 + a^2 p^2\right)+ \frac{m^2}{p^2} -
        \frac{m^4}{p^4}\,\log\left(1+\frac{p^2}{m^2}\right)\right)
        \Bigg]\Bigg] \nonumber \\
&& + \, \delta(p_1 - p_2)\, m \nonumber\\
&&\quad\Bigg[ 1 +
        \frac{g^2\,C_F}{16 \pi^2}\,\Bigg[ e_2 +5.79201\,\alpha
- (3+\alpha)\left( \log\left(a^2 m^2+
        a^2 p^2\right)+ \frac{m^2}{p^2}\,\log\left(1+\frac{p^2}{m^2}\right)\right)\Bigg]\Bigg],\,\,\,\,
\label{Sinverse}
\eea
where 
$p_1,p_2$: external momenta,  $a\,p_\rho\equiv\left(a\,p_{1\rho}+\pi/2\right)_{{\rm mod}\pi} - \pi/2 = \left(a\,p_{2\rho}+\pi/2\right)_{{\rm mod}\pi} - \pi/2\,$ and $C_F \equiv (N_c^2-1)/(2N_c)$\,.
The quantities $e_1,\,e_2$ are numerical coefficients that depend on the stout smearing parameters.
We have evaluated $e_1,\,e_2$ for several choices of 
Symanzik Improved gluon actions\,; in particular, for the Tree-Level Improved Symanzik gauge action we obtain:
\bea
e_1 = &-&7.2136+ 124.5148\,(\omega_{A_{1}}+\omega_{A_{2}})
- 518.4332\,(\omega_{A_{1}}^2+\omega_{A_{2}}^2)
- 2073.7329 \,\omega_{A_{1}}\,\omega_{A_{2}}\nonumber\\
&+& 9435.3458\,(\omega_{A_{1}}^2 \omega_{A_{2}} +\omega_{A_{1}}\,\omega_{A_{2}}^2)
- 45903.1373 \,\omega_{A_{1}}^2 \omega_{A_{2}}^2\,,\\
\label{e1}
e_2 = &&27.1081
-264.6953\,(\omega_{A_{1}}+\omega_{A_{2}})
+ 885.2150\,(\omega_{A_{1}}^2+\omega_{A_{2}}^2)
+3540.8600\,\omega_{A_{1}}\,\omega_{A_{2}}\nonumber\\
&-& 13960.0107(\omega_{A_{1}}^2 \omega_{A_{2}} + \omega_{A_{1}}\,\omega_{A_{2}}^2)
+ 60910.8220 \,\omega_{A_{1}}^2 \omega_{A_{2}}^2\,.
\label{e2}
\eea
In all numerical results (here and below) the systematic
error (coming from an extrapolation to infinite lattice size of our
numerical loop-integrals) is smaller than the last digit we present.

We denote the expression in square brackets, in the last line of Eq.~(\ref{Sinverse}), as
$\Sigma_m(p^2,m)$; from this we will extract the multiplicative
renormalization of the Lagrangian mass, $\Zm$. For bilinear operators ${\cal O}_\Gamma$, 
the RI$'$ renormalization scheme consists in imposing that the renormalized forward
amputated two-point Green's function, computed in the chiral
limit and at a given (large Euclidean) scale $p^2=\mu^2$, be equal to
its tree-level value. The RFs are computed at arbitrary values of the
renormalization scale $\mu$~\cite{Haris:2013}. 
Thus, the RI$'$ conditions for $\Zq$, $\Zm$, $Z_{{\cal O}_\Gamma}$ read:
\begin{eqnarray}
S_{1-loop}^{-1}(p^2=\mu^2,m=0)  &=& S_{tree}^{-1}(p^2=\mu^2,m=0) \Zq^{{\rm RI}'}(\mu)\\
 \Sigma_m(p^2=\mu^2,m=0) &=&\Zm^{{\rm RI}'}(\mu) \, \Zq^{{\rm RI}'}(\mu)\\
\Lambda_{{\cal O}_\Gamma}^{1-loop}(p^2=\mu^2,m=0) &=& \Lambda_{{\cal O}_\Gamma}^{tree}
\Zq^{{\rm RI}'}(\mu)\,\left(Z^{{\rm RI}'}_{{\cal
    O}_\Gamma}(\mu)\right)^{-1}\,, \quad (\Gamma=S,\,T,\,P)
\label{renormalization cond}
\end{eqnarray}
where $S_{tree}^{-1}$ is the tree-level result for the inverse
propagator, $\Lambda_{{\cal O}_\Gamma}^{tree}$ is the tree-level
value of the Green's function for ${\cal O}_\Gamma$ and $\Lambda_{{\cal O}_\Gamma}^{1-loop}$ is the {\it bare}
corresponding Green's function, up to one loop. The one-loop
Green's functions for ${\cal O}_V$ and ${\cal O}_A$ contain two
Lorentz structures each:
\be
\Lambda_{{\cal O}_V}^{1-loop} = \gamma_\mu \,\Sigma^{(1)}_V(p)+ \frac{p_\mu\,\pslash}{p^2} \,\Sigma^{(2)}_V(p)\,\,,\,\,\,\,
\Lambda_{{\cal O}_A}^{1-loop} = \gamma_5 \,\gamma_\mu \,\Sigma^{(1)}_A(p)+ \gamma_5 \frac{p_\mu\,\pslash}{p^2} \,\Sigma^{(2)}_A(p),
\ee
($\Sigma^{(2)}_{V,A}(p)= {{\cal O} (g^2)}$).
The presence of
$\Sigma^{(2)}_V$ and $\Sigma^{(2)}_A$  makes a prescription such as
Eq.~(\ref{renormalization cond}) inapplicable in those cases. Instead we apply renormalization 
conditions only on $\Sigma^{(1)}_{V,A}$:
\be
 \gamma_\mu \,\Sigma^{(1)}_V(p^2=\mu^2,m=0)\, =\, 
\Lambda_{{\cal O}_V}^{tree}\,
\Zq^{{\rm RI}'}(\mu)\,\left(Z^{{\rm RI}'}_{{\cal O}_V}(\mu)\right)^{-1}\,,
\label{VArenormCondition}
\ee
(and similarly for ${\cal O}_A$). 
We have also applied two stout-smearing steps to the links in the definition of bilinears,
with stout parameters $\omega_{{\cal O}_{1}}$ and $\omega_{{\cal O}_{2}}$; for general applicability all
stout parameters ($\omega_{A_{1}},\,\omega_{A_{2}},\,\omega_{{\cal O}_{1}},\,\omega_{{\cal O}_{2}}$) have been
kept distinct.

Our results for $\Zq$ and $Z_{{\cal O}_\Gamma}$ are presented below; we note that $\Zm$ and $\ZS$ turn out to
be related by $\Zm = \ZS^{-1}$, as was expected.
\begin{eqnarray}
 Z_q^{{\rm RI}'} = 1 + \frac{g^2\,C_F}{16 \pi^2}\,\Bigl[\hspace{-0.1cm}
&-&\hspace{-0.1cm} 7.2136 + 4.7920\,\alpha\, + 124.5149\,\left({\omega_{A_{1}}} + {\omega_{A_{2}}}\right)
- 518.4332\,\left({\omega_{A_{1}}}^2 + {\omega_{A_{2}}}^2 \right)\nonumber\\
&-& 2073.7329\,{\omega_{A_{1}}}\,{\omega_{A_{2}}}  
+ 9435.3459\,\left({\omega_{A_{1}}}^2\,{\omega_{A_{2}}}
+ {\omega_{A_{1}}}\,{\omega_{A_{2}}}^2\right)\nonumber\\
&-& 45903.1373\,{\omega_{A_{1}}}^2\,{\omega_{A_{2}}}^2 
+ \alpha \,\log\left(a^2\,\mu^2\right) \Bigr]
\end{eqnarray}  
\begin{eqnarray}
Z_S^{{\rm RI}'} =1 + \frac{g^2\,C_F}{16 \pi^2}\,
\Bigl[\hspace{-0.1cm}&-&\hspace{-0.1cm}
34.3217 -\,\alpha\, + 389.2102 \,\left({\omega_{A_{1}}} + {\omega_{A_{2}}}\right) 
- 1403.6482\,\left({\omega_{A_{1}}}^2 + {\omega_{A_{2}}}^2 \right) \nonumber\\
&-& 5614.5930\,{\omega_{A_{1}}}\,{\omega_{A_{2}}}
+ 23395.3566\,\left({\omega_{A_{1}}}^2\,{\omega_{A_{2}}} + {\omega_{A_{1}}}\,{\omega_{A_{2}}}^2\right)\nonumber\\
&-& 106813.9602 \,{\omega_{A_{1}}}^2\,{\omega_{A_{2}}}^2 
+ 3\,\log\left(a^2\,\mu^2\right) \Bigr]\,=\,(Z_m^{{\rm RI}'})^{-1}
\end{eqnarray}

\begin{eqnarray}
Z_P^{{\rm RI}'} = 1 + \frac{g^2\,C_F}{16 \pi^2}\,\Bigl[\hspace{-0.1cm}&&\hspace{-0.1cm}
25.7425 - \,\alpha\, + 119.0620\,\left({\omega_{A_{1}}} + {\omega_{A_{2}}}\right)
- 428.1202\,\left({\omega_{{\cal O}_{1}}} + {\omega_{{\cal O}_{2}}}\right)\nonumber\\
&-& 518.5414\,\left({\omega_{A_{1}}}^2 + {\omega_{A_{2}}}^2 \right) \nonumber\\
&+& 1667.0015\,\left({\omega_{{\cal O}_{1}}}^2 + {\omega_{{\cal O}_{2}}}^2 \right)
- 2042.2891\,{\omega_{A_{1}}}\,{\omega_{A_{2}}}
+ 6699.8826\,{\omega_{{\cal O}_{1}}}\,{\omega_{{\cal O}_{2}}}\nonumber\\
&+& 31.8765\,\left({\omega_{A_{1}}} + {\omega_{A_{2}}}\right)\,\left({\omega_{{\cal O}_{1}}} + {\omega_{{\cal O}_{2}}}\right)
+ 9435.3986\,\left({\omega_{A_{1}}}^2\,{\omega_{A_{2}}} + {\omega_{A_{1}}}\,{\omega_{A_{2}}}^2\right)\nonumber\\
&-& 29653.9826\,\left({\omega_{{\cal O}_{1}}}^2\,{\omega_{{\cal O}_{2}}} + {\omega_{{\cal O}_{1}}}\,{\omega_{{\cal O}_{2}}}^2\right)
- 210.2738\,\left({\omega_{A_{1}}} + {\omega_{A_{2}}}\right){\omega_{{\cal O}_{1}}}\,{\omega_{{\cal O}_{2}}}\nonumber\\
&-& 210.2738\,{\omega_{A_{1}}}\,{\omega_{A_{2}}}\,\left({\omega_{{\cal O}_{1}}} + {\omega_{{\cal O}_{2}}}\right)
- 44803.9568\,{\omega_{A_{1}}}^2\,{\omega_{A_{2}}}^2     \nonumber\\
&+& 143482.2565 \,{\omega_{{\cal O}_{1}}}^2\,{\omega_{{\cal O}_{2}}}^2     
+ 1657.7660 \,{\omega_{A_{1}}}\,{\omega_{A_{2}}}{\omega_{{\cal O}_{1}}}\,{\omega_{{\cal O}_{2}}} \nonumber\\
&+&3\,\log\left(a^2\,\mu^2\right) \Bigr]
\end{eqnarray}

\begin{eqnarray}
Z_V^{{\rm RI}'} 
= 1 + \frac{g^2\,C_F}{16 \pi^2}\,\Bigl[\hspace{-0.1cm}&&\hspace{-0.1cm}
86.7568\,\left[\left({\omega_{A_{1}}} + {\omega_{A_{2}}}\right)\,-\,\left({\omega_{{\cal O}_{1}}} + {\omega_{{\cal O}_{2}}}\right)\right]
- 337.3834\,\big[\left({\omega_{A_{1}}}^2 + {\omega_{A_{2}}}^2 \right)\,\nonumber\\
&-&\,\left({\omega_{{\cal O}_{1}}}^2 + {\omega_{{\cal O}_{2}}}^2 \right)\big]
- 1349.5337\,\left({\omega_{A_{1}}}\,{\omega_{A_{2}}}\,-\,{\omega_{{\cal O}_{1}}}\,{\omega_{{\cal O}_{2}}}\right)\nonumber\\
&+& 5950.8059\,\left[\left({\omega_{A_{1}}}^2\,{\omega_{A_{2}}} + {\omega_{A_{1}}}\,{\omega_{A_{2}}}^2\right)\,-\,\left({\omega_{{\cal O}_{1}}}^2\,{\omega_{{\cal O}_{2}}} + {\omega_{{\cal O}_{1}}}\,{\omega_{{\cal O}_{2}}}^2\right)\right]\nonumber\\
&-& 28627.2520\,\left({\omega_{A_{1}}}^2\,{\omega_{A_{2}}}^2\,-\,{\omega_{{\cal O}_{1}}}^2\,{\omega_{{\cal O}_{2}}}^2 \right)
 \Bigr]
\label{ZV}
\end{eqnarray}
\begin{eqnarray}
Z_A^{\rm RI'} = 1 + \frac{g^2\,C_F}{16 \pi^2}\,\Bigl[\hspace{-0.1cm}&&\hspace{-0.1cm}
17.0363 + 117.5844\,\left({\omega_{A_{1}}} + {\omega_{A_{2}}}\right)
- 314.3549\,\left({\omega_{{\cal O}_{1}}} + {\omega_{{\cal O}_{2}}}\right)\nonumber\\
&-& 518.4189\,\left({\omega_{A_{1}}}^2 + {\omega_{A_{2}}}^2 \right) \nonumber\\
&+& 1223.7950\,\left({\omega_{{\cal O}_{1}}}^2 + {\omega_{{\cal O}_{2}}}^2 \right)
- 2041.7996\,{\omega_{A_{1}}}\,{\omega_{A_{2}}}
+ 4927.0558\,{\omega_{{\cal O}_{1}}}\,{\omega_{{\cal O}_{2}}}\nonumber\\
&+& 31.8758\,\left({\omega_{A_{1}}} + {\omega_{A_{2}}}\right)\,\left({\omega_{{\cal O}_{1}}} + {\omega_{{\cal O}_{2}}}\right)
+ 9559.9779\,\left({\omega_{A_{1}}}^2\,{\omega_{A_{2}}} + {\omega_{A_{1}}}\,{\omega_{A_{2}}}^2\right)\nonumber\\
&-& 21823.5425\,\left({\omega_{{\cal O}_{1}}}^2\,{\omega_{{\cal O}_{2}}} + {\omega_{{\cal O}_{1}}}\,{\omega_{{\cal O}_{2}}}^2\right)
- 210.2735\,\left({\omega_{A_{1}}} + {\omega_{A_{2}}}\right){\omega_{{\cal O}_{1}}}\,{\omega_{{\cal O}_{2}}}\nonumber\\
&-& 210.2735\,{\omega_{A_{1}}}\,{\omega_{A_{2}}}\,\left({\omega_{{\cal O}_{1}}} + {\omega_{{\cal O}_{2}}}\right)
- 47154.2203\,{\omega_{A_{1}}}^2\,{\omega_{A_{2}}}^2     \nonumber\\
&+& 105753.7547 \,{\omega_{{\cal O}_{1}}}^2\,{\omega_{{\cal O}_{2}}}^2     
+ 1396.9376 \,{\omega_{A_{1}}}\,{\omega_{A_{2}}}{\omega_{{\cal O}_{1}}}\,{\omega_{{\cal O}_{2}}} 
\Bigr]
\end{eqnarray}
\begin{eqnarray}
Z_T^{{\rm RI}'} = 1 + \frac{g^2\,C_F}{16 \pi^2}\,\Bigl[\hspace{-0.1cm}&&\hspace{-0.1cm}
8.8834 + \,\alpha\, + 116.5787\,\left({\omega_{A_{1}}} + {\omega_{A_{2}}}\right)
- 200.5879\,\left({\omega_{{\cal O}_{1}}} + {\omega_{{\cal O}_{2}}}\right)\nonumber\\
&-& 531.7591\,\left({\omega_{A_{1}}}^2 + {\omega_{A_{2}}}^2 \right) 
+ 780.5904\,\left({\omega_{{\cal O}_{1}}}^2 + {\omega_{{\cal O}_{2}}}^2 \right)\nonumber\\
&-& 2095.1622\,{\omega_{A_{1}}}\,{\omega_{A_{2}}}
+ 3154.2357\,{\omega_{{\cal O}_{1}}}\,{\omega_{{\cal O}_{2}}}
+ 31.8743\,\left({\omega_{A_{1}}} + {\omega_{A_{2}}}\right)\,\left({\omega_{{\cal O}_{1}}} + {\omega_{{\cal O}_{2}}}\right)\nonumber\\
&+& 9877.2330\,\left({\omega_{A_{1}}}^2\,{\omega_{A_{2}}} + {\omega_{A_{1}}}\,{\omega_{A_{2}}}^2\right)
- 13993.1045\,\left({\omega_{{\cal O}_{1}}}^2\,{\omega_{{\cal O}_{2}}} + {\omega_{{\cal O}_{1}}}\,{\omega_{{\cal O}_{2}}}^2\right)\nonumber\\
&-& 284.0013\,\big[\left({\omega_{A_{1}}} + {\omega_{A_{2}}}\right){\omega_{{\cal O}_{1}}}\,{\omega_{{\cal O}_{2}}}+\,{\omega_{A_{1}}}\,{\omega_{A_{2}}}\,\left({\omega_{{\cal O}_{1}}} + {\omega_{{\cal O}_{2}}}\right)\big]\nonumber\\
&-& 48519.2862\,{\omega_{A_{1}}}^2\,{\omega_{A_{2}}}^2     
+ 68237.1178 \,{\omega_{{\cal O}_{1}}}^2\,{\omega_{{\cal O}_{2}}}^2 \nonumber\\    
&+& 2709.4942 \,{\omega_{A_{1}}}\,{\omega_{A_{2}}}{\omega_{{\cal O}_{1}}}\,{\omega_{{\cal O}_{2}}} 
-\,\log\left(a^2\,\mu^2\right) \Bigr].
\end{eqnarray}

We also note in passing that in the absence of stout smearing
($\omega_{A_i}=\omega_{{\cal O}_i}=0$)  $\ZV^{{\rm RI}'}=1$, as is well known from current
conservation. In addition, Eq.~(\ref{ZV}) shows that
non-renormalization of ${\cal O}_V$ applies also when
$\omega_{A_i}=\omega_{{\cal O}_i}$; this follows from the fact that
the stout link version of ${\cal O}_V$ mimics the fermion action, and
thus current conservation applies equally well in this case.

To obtain $Z_{{\cal O}_\Gamma}$ in the $\overline{\rm MS}$ scheme, we apply conversion factors, $C_{{\cal O}_\Gamma}$, which have been computed in dimensional regularization ~\cite{Gracey:2003yr}.
\be
Z_{{\cal O}_\Gamma}^{\overline{\rm MS}} = C_{{\cal O}_\Gamma}\,Z_{{\cal O}_\Gamma}^{{\rm RI}'}\,.
\label{Conv}
\ee
These conversion factors do not depend on the regularization scheme.
Furthermore, they refer to the Naive Dimensional
Regularization (NDR) of the $\overline{\rm MS}$ scheme, in which $C_P=C_S$ and $C_A=C_V$. From
Eq.~(\ref{Conv}) one obtains:
\be
\Zq^{\overline{\rm MS}}=\Zq^{{\rm RI}'} -
\frac{g^2\,C_F}{16\pi^2}\,\alpha,\,\,\,\,\,\,
Z_{\rm S,P}^{\overline{\rm MS}}=Z_{\rm S,P}^{{\rm RI}'} +
\frac{g^2\,C_F}{16\pi^2}\,(4+\alpha),\,\,\,\,\,\,
Z_{\rm V,A}^{\overline{\rm MS}}=Z_{\rm V,A}^{{\rm RI}'},\,\,\,\,\,\,\ZT^{\overline{\rm MS}}=\ZT^{{\rm RI}'} -
\frac{g^2\,C_F}{16\pi^2}\,\alpha.\,
\ee

\section{Simulation setup}

We study the effect of an external magnetic field, B, on
the expectation value of the tensor polarization~\cite{Bali:2012jv}. 
The leading order of the expectation value of the tensor polarization operator $\bar{\psi}_f \sigma_{\mu\nu} \psi_f$, is proportional to the field strength and thus can be written as:
\be
\expv{ \bar\psi_f \sigma_{xy} \psi_f } = q_f B \cdot \expv{\bar \psi_f\psi_f}\cdot \chiF_f \equiv q_f B \cdot \TAU,
\label{eq:defchi}
\ee
where $\expv{\bar \psi_f\psi_f}$ is the quark condensate and $\chiF_f$ is the ``magnetic susceptibility'' of the condensate. We define the ``tensor coefficient'' $\TAU$ as the product of the condensate and the magnetic susceptibility of the condensate.

The bare observable can be written as:
\be
\expv{\bar\psi_f\psi_f}(B,T) = \frac{1}{\ZS}\expv{\bar\psi_f\psi_f}_{\rm renorm}(B,T) + \zeta_S m_f/a^2+\ldots,
\label{eq:renS}
\ee
where the divergences in $\expv{\bar \psi_f\psi_f}$ depend neither on the temperature nor on the external field. Therefore, in mass-independent renormalization schemes, $\zeta_S$ and $\ZS$ are just functions of the gauge coupling. Similarly,
\be
\expv{\bar\psi_f\sigma_{\mu\nu}\psi_f}(B,T) =\frac{1}{\ZT} \expv{\bar\psi_f\sigma_{xy}\psi_f}_{\rm renorm}(B,T) + \zeta_T q_fB m_f\log(m_f^2a^2)+\ldots,
\label{eq:renT}
\ee
where $\zeta_T$ is the coefficient of the divergent logarithm. Both $Z_T$ and $\zeta_T$ are independent of $T$ and $B$ (and $m_f$, in mass-independent schemes). In Eq.~({\ref{eq:renT}}) the ellipses denote finite terms. In the free theory we calculate $\zeta_T(g=0)=3/(4\pi^2)$. We use the perturbative $Z_S$ and $Z_T$ in $\overline{\rm MS}$ scheme with parameters ${\omega_{{\cal O}_{i}}}={\omega_{A_{i}}}=0.15$. We also notice that the operator $1-m_f\partial/\partial m_f$ eliminates the logarithmic divergence and thus can be used to define an observable with a finite continuum limit,
\be
\TAU^{\rm renorm} \equiv\TAU^r = \left(1-m_f\frac{\partial}{\partial m_f}\right)\TAU \cdot \ZT \equiv \TAU \ZT - \TAU^{\rm div},
\label{eq:mdmsub}
\ee
where $\TAU^{\rm div}$ is independent of the temperature. We measure  $Z_T \cdot \tau_f$, with the value of $m_s$ fixed to its physical value and we tune only $m_{ud}$ ($R\equiv m_{ud}/m_{ud}^{\rm phys}$) for $T=0$ and $0.5 < R < 28.15$. We consider the following fit function for $Z_T \cdot \tau_f$: 
\be
c_{f0} + c_{f1} R + c_{f2} R \log(R^2a^2),\quad 
 c_{fi}=c_{fi}^{(0)}+c_{fi}^{(1)}a^2
\ee
Appling the operator $1-m_f\partial_{m_f}=1-R\partial_R$ we find the value of $\tau^r_f$ at $T=0$.
We translate these results to the magnetic susceptibility $\chiF_f$ of Eq.~(\ref{eq:defchi}) using the value of the quark condensate. The zero-temperature magnetic susceptibilities (in the $\MSBar$\,scheme at scale $\mu=2 \textmd{GeV}$) are:
\be
\chiF_u = - (2.08\pm0.08) \textmd{ GeV}^{-2},\,\,\,\,\,\,\chiF_d = - (2.02\pm0.09) \textmd{ GeV}^{-2},\,\,\,\,\,\,\chiF_s = - (3.4\pm1.4) \textmd{ GeV}^{-2}.
\label{eq:chiresult}
\ee
We observe that the dependence of the condensate on $B$ varies strongly with the temperature in the transition region.
More details on the simulation setup and nonperturbative results can be found in  Refs.~\cite{Bali:2012jv, Endrodi:2012}.

\section{Conclusion}

We have determined the renormalization functions of taste-singlet bilinear fermion operators.
The novelty in our perturbative calculations is the stout smearing of the links
that we apply in both the fermion action and in the bilinear operators.
More precisely, we use two steps of stout smearing with
distinct parameters. To make our results as general as possible
we also distinguish between the stout parameters appearing in the fermion
action and in the bilinears.

In Ref.~\cite{Bali:2012jv, Endrodi:2012}, we measure the magnetic susceptibilities $\chi_f$ at zero temperature for the up, down and strange quarks in the $\MSBar$ scheme at a renormalization scale of $2 \textmd{ GeV}$.
The magnetic susceptibilities at $T=0$ are negative, indicating the spin-diamagnetic nature of the QCD vacuum.
We also find that the polarization changes smoothly with temperature in the confinement
phase and then drastically reduces around the transition region.
\bigskip

{\bf Acknowledgements:} Our work was supported by the Cyprus Research Promotion Foundation under Contract No. TECHNOLOGY/$\Theta E\Pi I\Sigma$/0311(BE)/16 and the EU (ITN STRONGnet 238353).

\end{document}